\documentclass[11pt,twoside]{article}

\usepackage{asp2004}
\usepackage{epsf}
\usepackage{lscape}
\usepackage{amssymb}

\newcommand{\mdot}{\ensuremath{\dot{M}}}                             % mass loss rate
\begin{document}
\title{Influence of the frictional heating on the wind line-profiles of SMC
stars}
\author{Ji\v{r}\'{\i} Krti\v{c}ka}   %%% Fill in author names
\affil{\'Ustav teoretick\'e fyziky a astrofyziky, P\v r\'\i rodov\v edeck\'a\
fakulta Masarykovy univerzity, Kotl\'a\v rsk\'a 2, CZ-611 37 Brno,
Czech Republic}    %%% Fill in author affiliations
\author{Daniela Kor\v c\'akov\'a and Ji\v{r}\'{\i} Kub\'at}
\affil{Astronomick\'y \'ustav, Akademie v\v{e}d
        \v{C}esk\'e republiky, CZ-251 65 Ond\v{r}ejov, Czech Republic}

\begin{abstract} %%% Abstract to run on from here.
We study the influence of the frictional heating on the wind line-profiles of
SMC stars. For this purpose we use our NLTE wind code to obtain consistent
occupation numbers of studied levels and our radiative transfer code to solve
the radiative transfer equation in moving media. We compare predicted wind line
profiles calculated with and without frictional heating for a~low-luminosity SMC
star with a weak wind and discuss the relevance of frictional heating as a
solution of a "weak-wind problem" for this star.
\end{abstract}

\section{Introduction}

\begin{figure}[h!]
{\centering%
\resizebox{0.36\hsize}{!}{\includegraphics{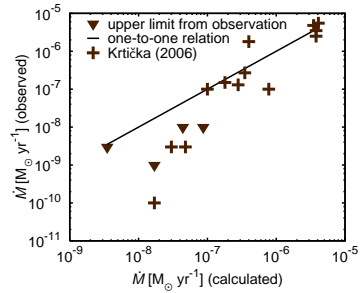}}\\}
\caption{Comparison of predicted \mdot\ and \mdot\ derived
from observations.}
\label{dmdt}
\end{figure}

For lower luminosity hot stars the predicted mass-loss rates \mdot\ are
significantly higher than those inferred from observations \citep[``weak-wind
problem'', e.g.][see also Fig.~\ref{dmdt}]{martin}. Wind structure of these
stars may be modified by the effects connected with multicomponent nature of
wind flow (like frictional heating). Here we discuss influence of frictional
heating on wind line profiles.

\section{Line profiles of four-component wind model of SMC N81 \#2}

For the calculation of multicomponent stationary spherically symmetric NLTE wind
model we apply the code by Krti\v{c}ka \& Kub\' at (\citeyear{nltei}). The
output from our wind model we insert into our radiative transfer model developed
by Kor\v{c}\'{a}kov\'{a} \& Kub\'{a}t (\citeyear{osa}) that solves the radiative
transfer equation in axial symmetry.

\begin{figure}[ht]
\plottwo{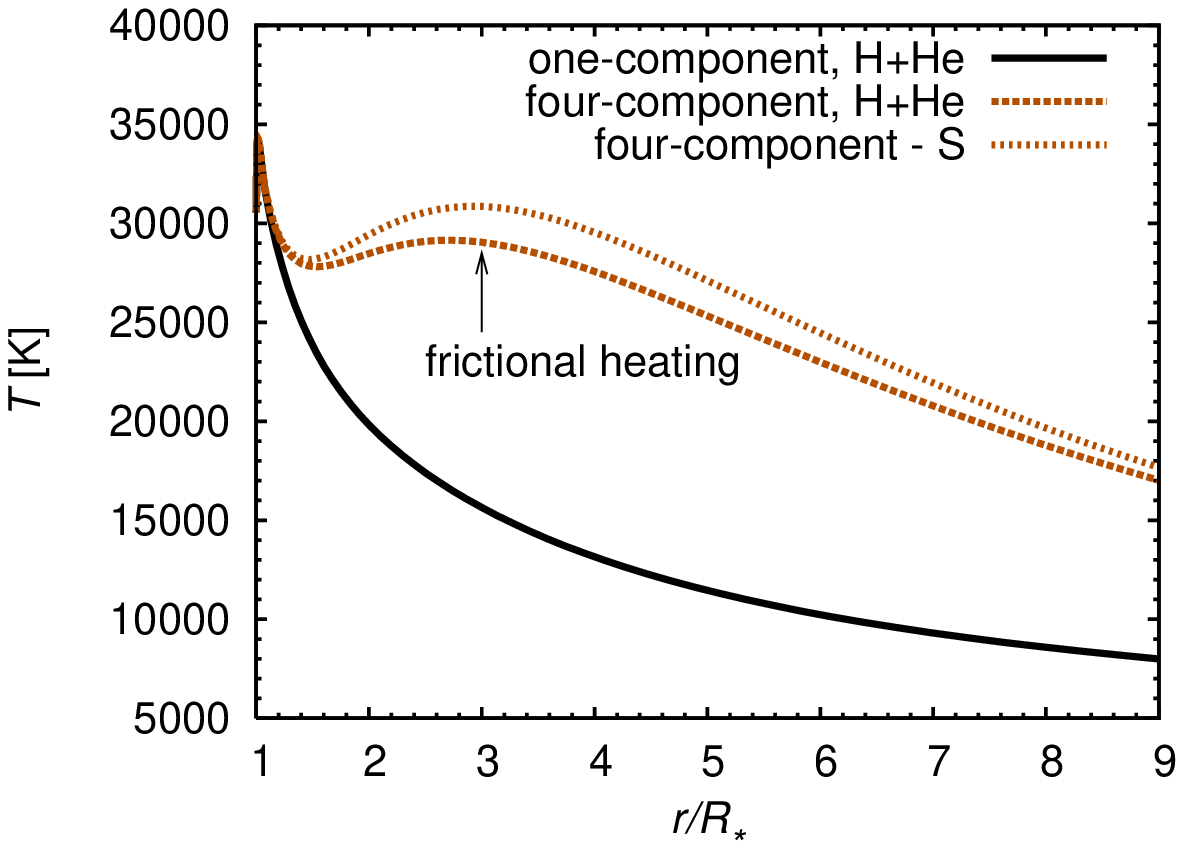}{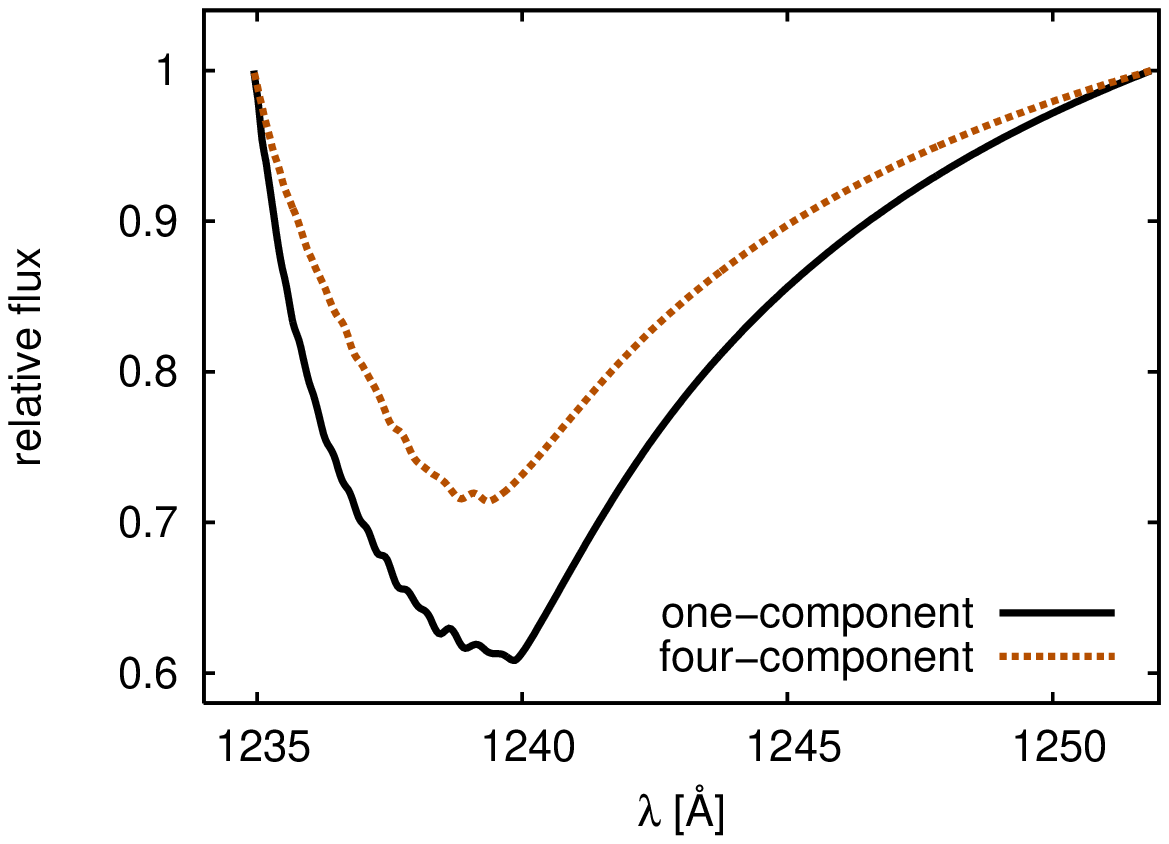}
\caption{Left panel: Temperature structure of four-component wind model
compared to the one-component one. Right panel: Comparison of
\ion{N}{v} 1242\,\AA\ line profiles calculated using a
model with and without frictional heating.}
\end{figure}

\citet{nlteii} calculated a multicomponent wind model of the star SMC N81~\#2
from the Small Magellanic Cloud (SMC) with parameters taken from \citet{martin}.
He explicitly included four wind components: passive~component (hydrogen,
helium), sulphur, other heavier elements, and free electrons.

Close to the stellar surface ($r/R_*\lesssim1.3$) the wind density is high,
consequently, the velocity difference between wind components is small and
multicomponent effects are negligible. With increasing velocity the
density decreases, and, as a consequence, the velocity difference between
wind components is higher. The velocity difference between sulphur and passive
component is so large that friction between these wind components significantly
influences the temperature. The wind temperature in the four-component
model is higher than in the one-component model (in which the effects of
frictional heating are not included).

The \ion{N}{v} 1242\,\AA\ line profile is plotted for both multicomponent
stellar wind (with frictional heating) and one-component model (without
frictional heating). The frictional heating modifies the ionization equilibrium.
Consequently, in the case of neglected frictional heating (a commonly used
simplification) the line profile is deeper than that calculated using model with
frictional heating. Since the "observed" mass-loss rate is obtained  by fitting
the wind line-profiles by appropriate model, its  value derived using models
with frictional heating is higher. Such value is closer to the predictions of
hydrodynamical calculations.

Our preliminary results show that part of the "weak-wind problem" solution can
be connected with improvement of wind ionization equilibrium models.

\acknowledgements
Grants GA \v{C}R 205/03/D020, P205/04/P224, and GA AV \v{C}R grant B301630501.

\end{document}